# Network analysis of protein dynamics

Csaba Böde [a,¶], István A. Kovács [b,¶], Máté S. Szalay [b,¶], Robin Palotai [b,¶],
Tamás Korcsmáros [b,¶] and Péter Csermely [b,*]

*Department of [a]Biophysics and Radiation Biology and [b]Medical Chemistry,
Semmelweis University, Puskin str. 9, H-1088 Budapest, Hungary*

The network paradigm is increasingly used to describe the topology and dynamics of complex systems. Here we review the results of the topological analysis of protein structures as molecular networks describing their small-world character, and the role of hubs and central network elements in governing enzyme activity, allosteric regulation, protein motor function, signal transduction and protein stability. We summarize available data how central network elements are enriched in active centers and ligand binding sites directing the dynamics of the entire protein. We assess the feasibility of conformational and energy networks to simplify the vast complexity of rugged energy landscapes and to predict protein folding and dynamics. Finally, we suggest that modular analysis, novel centrality measures, hierarchical representation of networks and the analysis of network dynamics will soon lead to an expansion of this field.

**1. Introduction: topological networks of protein structures**

The network concept is widely used to analyze and predict the dynamics of complex systems. When talking about networks, the complex system is perceived as a set of interacting elements (nodes, vertices), which are bound together by links (contacts, edges, interactions). In usual networks (graphs) links represent interactions between element pairs. Links usually have a weight, which characterizes their strength (affinity, intensity or probability). Links may also be directed, when one of the elements has a larger influence to the other than vice versa. Most self-organized networks are small worlds, where two elements of the network are separated by only a few other elements. Networks contain hubs, i.e. elements, which have a high degree (or in other words: have a large number of neighbors). Random networks have a Poissonian degree distribution, which means that they have a negligible amount of hubs. On the contrary, in many networks we observe a scale-free degree distribution, which means that the probability to find a hub with a number of neighbors a magnitude higher is a magnitude lower (but, importantly, not negligible). Networks can be dissected to overlapping modules (communities, groups), which often form a hierarchical structure [1-7].

We must warn that the above summary of the major features of self-organizing, real-world networks is largely a generalization, which is often not observed in its pure form. Real world networks are often heterogeneous, and their different modules may behave completely differently. Moreover, sampling bias and improper data analysis may show the above features in such cases, where they do not actually exist. Therefore, special caution has to be taken to scrutinize the validity and extent of datasets, use correct sampling procedures and adequate methods of data analysis [8-11].

[¶]Csaba Böde (csabi@puskin.sote.hu), István A. Kovács (steve3281@bolyai1.elte.hu), Máté S. Szalay (szalaymate@gmail.com), Robin Palotai (palotai.robin@gmail.com) and Tamás Korcsmáros (korcsmaros@gmail.com) started their research as members of the Hungarian Research Student Association (www.kutdiak.hu), which provides research opportunities for talented high school students since 1996.

*\*Correspondence to:* Péter Csermely, Department of Medical Chemistry, Semmelweis University School of Medicine, Budapest, P.O. Box 260. H-1444 Hungary. Telephone: +361-266-2755 extn.: 4102. fax: +361-266-6550. E-mail: csermely@puskin.sote.hu

**Table 1. Protein structure, energy and conformational networks**

| Definition of links in the network[a] | Usual purpose of network representation | References |
|---|---|---|
| If the distance between amino acid side chains is below a cut-off distance (usually between 0.45 and 0.85 nm) → un-weighted link | Detection of details in protein structure | [12,16] |
| Distance between βC atoms → weighted link | Detection of details in protein structure | [14] |
| Weight is constructed from the number of possible links between the two amino acid side chains, if the distance between amino acid side chains is below a cut-off distance → weighted link | Detection of details in protein structure | [25] |
| Hydrogen bonds | Analysis of protein structure and dynamics | [24] |
| Distance between αC atoms → weighted link treated as a spring | Construction of an elastic network model to assess protein dynamics | [41] |
| Treat all distances as spring and form a spring network | Construction of an elastic network model to assess protein dynamics | [43] |
| Conformational transitions | Predict native structure and assess the probability of conformational transitions | [27,63,64] |
| Saddles of the energy landscape (representing conformational transitions) → un-weighted or weighted links | Simplify the multitude of basins on rugged energy landscapes to predict protein folding pathways | [55,57,59] |

[a]Protein structure networks are also called as amino-acid networks, residue-networks or protein structure graphs to discriminate them form 'protein networks', which is a widely used term for protein-protein interaction networks.

## 2. Topological networks of protein structures

In protein structure networks network elements represent segments of the protein, while their weighted links are constructed by taking into account the physical distance between these elements. Network elements can be atoms, like the αC or βC atoms of amino acids. However, most of the times elements of protein structure networks are whole amino acid side chains. Currently, un-weighted protein structure networks are much widely used than weighted ones. In un-weighted protein structure networks a cut-off distance (which is usually between 0.45 and 0.85 nm, Table 1.) is introduced, and only those amino acid side chains are connected with un-weighted links, which are nearer to each other than the threshold set by the cut-off distance (Fig. 1.). These networks are usually called amino-acid networks, residue-networks or protein structure graphs to discriminate them from 'protein networks', which is a widely used term for protein-protein interaction networks. We will use the term 'protein structure network' in this paper to denote this type of description of protein-residue topology. Protein structure networks have been used first as a form of data-mining to help the structure comparison of proteins and to identify structural similarities [12,13]. However, after 1998 the approach started to use the expanding knowledge of network studies, which led to several important results, which we will describe in detail in the following sections [14-16].

As an exception from most self-organized networks, the degree distribution of protein structure networks seems to be Poissonian and not scale-free [17,18]. The Poissonian degree distribution means that protein structures have a much smaller number of hubs than most self-organized networks including most cellular or social networks. The major reason for this deviation from the scale-free degree distribution lies in the limited simultaneous binding capacity of a given amino acid side-chain (also called as excluded volume effect). The explanation behind the scale-free degree distribution of macromolecular assemblies is that macromolecules (e.g. proteins) have much less constraints to increase their contact surface, and are not restricted to simultaneous binding only, since they may leave their partners and bind to different neighbors. Similar assumptions (to a greater extent) hold to us while forming social networks.

The limited amino acid side chain binding capacity contributes to the fact that each amino acid has a characteristic average degree. This depends on the interaction cut-off, which makes hydrophilic amino acids 'strong hubs' (observed at high interaction cut-off allowing low overlaps), and hydrophobic amino acids 'weak hubs' (at low interaction cut-off allowing high overlaps), respectively. Hubs are integrating various secondary structure elements, and, therefore, it is not surprising that they increase the thermodynamic stability of proteins [19,20].

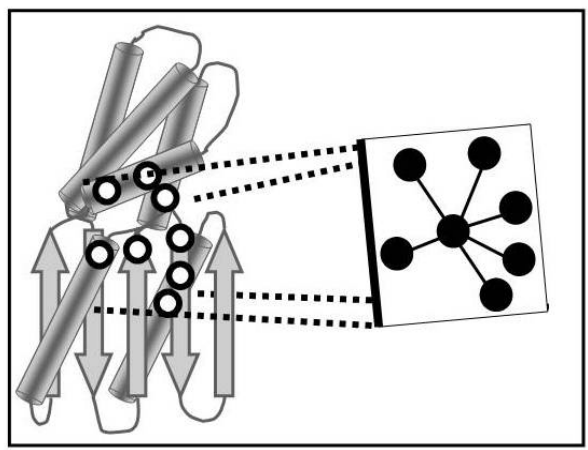

**Fig. 1. A protein structural network.** An illustrative segment of a protein structural network (right) is derived from a 3D representation of a protein (left), where distinct parts (atoms or most of the times whole amino acid side chains, open circles on the left) will be the network elements (black filled circles on the right), while the links of the network (solid lines on the right) are constructed by taking into account the physical distance of the respective protein parts from each other. Please note, that in a more detailed picture these topological links can also be strong and weak depending on distance (correlating in many cases with the bond-strength within certain limits) between the respective protein segments.

Key amino acids (nucleation centers), which were shown to govern the folding process, are central residues of the topological network representing the transitional conformation. However, central amino acids of the transitional conformation are not the same as central amino acids of the native conformation reflecting a gross-rearrangement of protein networks during the folding process [14,16,21]. Similarly, a redistribution of central residues was observed, when active and inactive conformations of hemoglobin were compared [22]. Residues with small average of their shortest path lengths (also characterized by the centrality measure of the inverse of the mean shortest path lengths, called closeness or inverse geodesic length) are often found in the active or ligand binding sites of proteins [23]. This may reflect that active or binding sites are preferentially centered within the protein structure network. Central amino acids have also been revealed by the analysis of hydrogen bonding networks (HB plots), i.e. 2D representations of hydrogen-bonds of non-adjacent amino acids [24].

Protein structure networks are assortative (meaning that their hubs preferentially associate with other hubs), and have a hierarchical structure (there are central hubs, which associate with more hubs and 'peripheral hubs', which have less hub neighbors than the central hubs). Interestingly, both the assortativity and hierarchical structure is valid only to the protein structure subnetwork of hydrophobic amino acids, but can not be observed with the subnetworks of hydrophilic and charged amino acids confirming the key role of hydrophobic interactions in the core-structure of proteins [25].

Proteins are small-worlds. In the small-world of protein structures any two amino acids are connected to each other via only a few other amino acids. This feature is true to most globular and fibrous proteins [17,18,26,27]. Small-worldness is valid to the protein residues residing both in the protein core and on the surface of proteins [17]. Dokholyan et al. [21] found that small-world type connectivity of the protein structure network determines folding probability (proteins with denser protein structure networks fold easier), and the small-worldness of the protein structure network increases during the folding process as the protein structure becomes more and more compact. However, we must warn that most observations above were based on un-weighted small-worlds. Assessment of weighted small-worlds may give interesting surprises in the future.

Motif (pattern) search in protein structure networks has also been addressed in detail. Motifs are widely and characteristically occurring assemblies of a few network elements (typically three to six amino acid side chains), which can be identified, if members of an evolutionary related protein set consisting five or more proteins are compared. Such motifs can be the well-known Ser/His/Asp catalytic triad, the zinc-finger or EF-hand metal coordination sites, etc. However, the number of 'meaningful' motifs is much higher than this, and can be in the range of 500 in a given protein set. Several network-based programs, such as ASSAM or DRESPAT have been developed for the search of motifs in protein structure networks [13,28].

Protein structure networks often have modules (i.e. communities of amino acids, which have a much higher intra-modular density, than the density of their inter-modular contacts linking them to other modules). These network modules have been determined by spectral graph-clustering methods of protein structural networks, and were shown to correspond to protein domains [14]. Domains tend to move together, which was used to dissect the inter-domain residues, which are important in regulation of protein function [29,30]. Locally dense structures of hydrogen-bond networks of proteins have been called as 'stabilization centers' and were identified with the program SCide [31].

Domains usually fold separately, have a function and are conserved during evolution. The distribution of the folds of various domains follows a scale-free pattern [32] meaning that there is a small number of very 'popular', stable folds, and we have a relatively big number of unique, orphan folds. The underlying reason of the 'popular' folds is evolutionary selection, which preferred those structures, which are both stable and fold easily. These structures are the ones, which have the common feature of the small-worldness and the other topological specialties, which were either mentioned above, or will be detailed further in Section 5.

### 3. Unstructured regions: a transition to protein dynamics

Unstructured proteins (or unstructured protein regions), which are also called as intrinsically disordered proteins (IDPs) became a focus of intensive studies in recent years. The lack of conventional secondary structure in protein segments or in entire proteins helps a lot of binding and recognition processes, and increases the dynamics of both single proteins and protein complexes [33]. However, the disorder of protein structure is a matter of time-scale, and is much more prevalent than it is thought to the first glance. Flexibility of the polypeptide-chain leads to structural fluctuations [34]. However, this 'short-term' disorder is caused by fluctuations around an equilibrium conformation, which is different from the lack of equilibrium conformation observed in unstructured protein regions. We will summarize these dynamical aspects of protein disorder in the next Section.

### 4. Protein dynamics: quasi-harmonic movements, restricted relaxation and avalanches

The early work of Ansari et al. [35] already showed the existence of 'protein-quakes', i.e. the cascading relaxation avalanche of myoglobin after the photodissociation of carbon monoxide. A number of protein kinetics, including the above mentioned carbon monoxide dissociation, enzyme actions, exchange of protein protons to those of water and protein folding, are similar to Levy-flights, and show a scale-free statistics in the time-gaps between elementary conformational changes as well as in the magnitude of these changes [36-38]. Scale-free distributions and avalanches resemble to the behavior in 'self-organized criticality', and are typical features of systems with restricted relaxation [4]. In proteins the restrictions come from the necessity to break bonds in large-scale conformational transitions, which can be called as a local unfolding event. However, most protein motions (such as those observed after ligand binding) do not require bond-rearrangements and can be well approximated by quasi-harmonic dynamic [39].

In most conformational rearrangements the above scale-free distributions become more complex, which is due to the hierarchical and modular structure of the underlying protein structural network. In these real scenarios we observe the integration of the correlated scale-free distributions of the individual, overlapping network modules [36-38]. As an example of the inter-modular correlation of protein dynamics, Balog et al. [29] recently showed that conformational transitions of the individual domains are not additive in the simulation of phosphoglycerate kinase dynamics. Correlated motions of a network of distant residues have also been observed in dihydrofolate reductase [40].

As an example for the use of protein structural networks for the analysis of protein dynamics, fluctuations of amino acid side chains are correlated with the mean of the shortest path lengths of the amino acid in the protein structural network [17]. This reflects that more central amino acids (having a shorter average of their shortest path lengths) have a more restricted motion. Protein structural networks take into account only the interactions between amino acid side-chains, and neglect the constraints of the protein backbone. This is not a problem, if we analyze the topology of these networks, and want to draw conclusions for the structure and stability of proteins. However, it may restrict the analysis, when we would like to use the dynamics of topological networks to explain protein motions and rearrangements. This problem is circumvented by the elastic network model, where only the atomic coordinates of the αC atoms are used to build the network. Here a harmonic potential is used to account for pairwise interactions between all αC atoms within a cut-off distance, which was 1 nm in the study of Zheng et al. [41].

Using the above elastic network analysis a set of sparsely connected, highly conserved residues were identified, which are key elements for the transmission of allosteric signals in three nanomachines, such as DNA polymerase, myosin and the GroEL chaperonin [41]. Importantly, central amino acid residues in 'conventional' protein structure networks were also identified as strategically positioned, highly conserved key elements of allosteric communication by other network constructing methods using both βC atoms, or whole amino acid side chains [14,22]. These agreements indicate that the above network construction methods (Table 1.) complement and support each other. Clusters of amino acids around the active centers or ligand binding sites expand in an unparalleled, unique fashion, if the cut-off distance is increased, which also shows the unique centrality of these key functional segments – now at a higher level of network structure [14]. In agreement with the above observation, protein motions of substrate-free enzymes were shown essentially the same as the characteristic motions during catalysis, and had a frequency corresponding to the catalytic turnover rate. These motions extend much beyond the active center, which here again implies that concerted motions of a wide network of residues spanning the entire protein help enzyme catalysis [42].

Another elastic network representation treats all atomic distances as springs, and forms a spring network (Table 1.). Using this approach overconstrained (having more crosslinking bonds than needed) and underconstrained (with less crosslinking bonds than needed) protein regions were identified. These regions were nicely corresponding with rigid and flexible protein segments, respectively [43].

Protein dynamics can also be assessed by analyzing the propagation of perturbations in the hydrogen-bond network of the protein. A simplified, 2D network representation of hydrogen bonds, called HB-plot already revealed a number of key features of protein dynamics in the examples of cytochrome P450 and ligand-gated ion channels [24]. Hydrogen-bond rearrangements are also key elements of the involvement of water in protein dynamics as described in the next Section.

### 5. Protein dynamics: Water as a lubricant
Proteins may also 'borrow' flexibility from their surrounding. Water helps to overcome many kinetically restricted segments of protein motion acting as a 'lubricant'. Water molecules make a hydrogen-bond network as well as fluctuating hydrogen bonds with peptide bonds and amino acid side chains [4,44-47]. These transient changes induce a fluctuation in the energy level of the actual protein conformation, and open a possibility for a transient decrease in the activation energy between various conformational states. In agreement with these assumptions, a paper from Peter Wolynes' lab [48] showed that water efficiently lowers the saddles (activation energies) of the energy landscapes and makes previously forbidden conformational transitions possible. Interestingly, water-induced fluctuations decrease as protein folding proceeds [49], which may indicate a decreased help for protein folding as the multitude of conformational states converge to the native conformation. The detailed analysis of the contribution of water molecules to the hydrogen-bond networks of proteins awaits further investigation.

We have quite numerous and sometimes contradictory observations on the residual protein mobility in the absence of water [4,45-47]. On one hand, a 'monolayer' of water molecules and their hydrogen-bond network is needed on the protein surface to restore the dynamics of biomolecules. The dynamics emerges, when the individual water molecules establish the percolation of their hydrogen-bond network [50]. On the other hand, in many enzymes a residual enzyme activity can still be observed at very low hydration levels [51]. Detailed investigations were able to discriminate protein movements, called slaved processes, which need the contribution of water as the solvent, and movements, which are independent of the solvent, called nonslaved processes [52]. Though several proteins can withstand a transfer to non-aqueous media, most enzymatic functions are stopped in the complete absence of water. Moreover, several dry proteins have a 'memory'. They preserve enzyme activity, if their structure has been previously stabilized. These dry proteins 'remember' to their active state, since their conformational changes are frozen in the absence of water [53]. Network analysis of hydrogen-bond networks at different hydration levels will be an exciting task of the future.

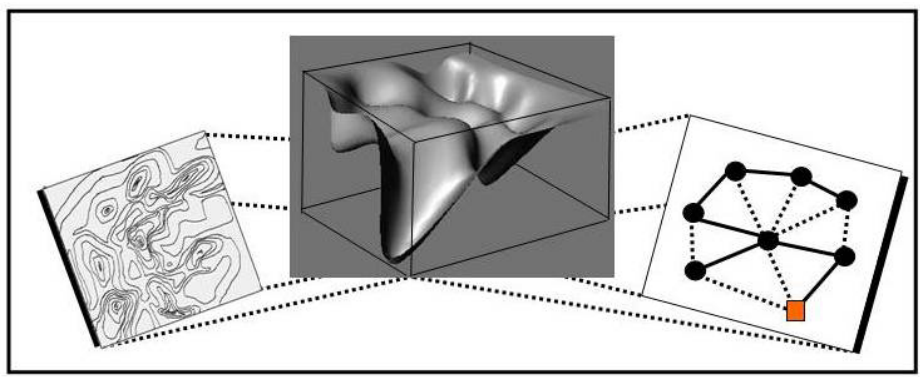

**Fig. 2. Energy network representation of the conformational transitions of protein dynamics.** An illustrative energy landscape is shown as a 3D image (center) and as a contour plot (left). On the right its transformation to an energy network is described. In the energy network representation (right) nodes represent local energy minima, while solid and dotted lines denote strong and weak links representing low and high activation energy transitions between two local energy minima, respectively. The rectangle on the bottom right of the network represents the lowest energy state to mark the native state of the respective protein.

## 6. Energy and conformational networks in the description of protein dynamics

Conformational states of proteins can be efficiently described by energy landscapes (Fig. 2.). The energy landscape may be simplified to an energy network. Here nodes of the network represent local energy minima and links between these energy minima correspond to the transition states (saddles) between them (Table 1.). The energy network of proteins has both a small-world and a scale-free character [54-56]. The assessment of weighted small-worlds will be a task of the future and may give interesting surprises. A weighted version of the energy network has been recently described by Gfeller et al. [57], where module determination methods were used to find the basins of the underlying energy landscape. This approach is helpful all the more, since the number local minima on the energy landscape is an exponential function of the residues involved [58], and requires a simpler, 'renormalized' representation to handle and understand its complexity both computationally and cognitively.

The modularized energy network proved to be heterogeneous, where scale-free-type degree distributions were observed only in that part of the modules, which had a major contribution of enthalpy changes (enthalpy-dominated energy basins of the underlying energy landscape). On the other hand, entropy-dominated modules showed a Gaussian degree-distribution pattern [57]. The restriction of scale-free degree distribution to network segments and the overlap of scale-free distribution with a Gaussian degree distribution agrees well with recent findings on topological networks [9-11]. The 'complexity' of energy networks (in this very rough sense meaning the number of energy basins on the energy landscape) has been suggested as an important measure of the 'ruggedness' of the energy landscape helping the discrimination between 'easy folder' proteins from those, which get stuck in the morass of possible conformations [59]. We have to note, that to define the links between network topology and complexity in the numerical sense (meaning e.g. the number of individual parameters necessary to predict the behavior of the network) is a very difficult task, which will be a potential breakthrough of the future.

Modularization of the energy network may also help us to solve the basic dilemma of the definition of energy networks, i.e. "What may we regard as a local energy minimum of the underlying energy landscape?" Local minima are by far not only sharp, well-defined topological features of the energy landscape. Many times local minima may form or may temporarily expand to shallow local basins with numerous fluctuating 'real' minima inside. Therefore, a more exact approach is to take all possible conformations as a 'local minimum' and determine the basins as primary modules of the resulting hierarchical networks.

Additionally, we may also think on the directedness of the energy networks. In principle, the higher is the difference between the energy of local neighboring energy minima, the more directed is the link between the two minima in the energy network.

The small-worldness of the energy network may give an underlying explanation of the high dynamism of protein structure: a node of the network representing a protein conformation is only a few steps (conformational transitions) apart from any other protein conformations. The energy landscape is hierarchical, and contains a number of hierarchically organized traps, which explain well the non-exponential, stretched kinetics in the early phase of protein folding as well as the aging of proteins at cryogenic temperatures [60-62]. This hierarchical nature makes the energy network resemble to a fractal-like structure, similar to that of the Apollonian networks [58].

Another network representation of the energy levels behind protein conformations is the 'conformational network' (also called configuration space network) of proteins, where the individual nodes are corresponding to the conformations, and the links are the conformational transitions between them (Table 1. [27,63,64]). The energy networks above and the conformational networks here obviously highly resemble to each other, since essentially they are representing the same ensemble of protein states – approaching it from different data-sets using slightly different rules. Both networks were used to predict the native protein structure as well as to assess the probability of various conformational transitions.

The combination of the 'conformational networks' (energy networks) with the underlying multitude of the respective protein structural networks of the individual protein conformations can be tackled by the analysis of the dynamics of protein structural networks. This important task will be a key development of future studies as we highlight in the next Section.

**7. Summary and perspectives**
In summary, we have shown that general assumptions of network studies, such as the small-world character and the scale-free degree distribution of many real-world networks had a great impact on our understanding of both protein structure networks and protein conformational/energy networks.

- Both protein structure networks and conformational networks are small worlds, which reflect the compactness and explain the exceptionally high dynamism of protein structure, respectively. Hydrophobic amino acids seem to play a more important role in the integration of protein structure networks than hydrophilic or charged amino acids, which shows the importance of the hydrophobic core of globular proteins.
- Hubs and central residues are integrating secondary structure elements, and increase protein stability. Central residues are strategically positioned, govern many conformational changes, and are often essential for the transduction of allosteric signals. Central residues are often found in the active, or ligand binding sites of proteins, and make these protein segments central parts of the topological organization of protein structure. This may explain why active centers and ligand binding sites often govern the dynamics of the entire protein triggering extreme avalanches of protein motions during enzyme catalysis or signal transduction.
- The modules (communities) of protein structure networks already helped us to identify key inter-modular residues, which often govern conformational transitions at domain boundaries. Modular analysis of conformational/energy networks is essential to simplify rugged energy landscapes 'renormalizing' them to a form, which is both computationally and cognitively tractable. This will help us both to discriminate between 'easy folder' proteins from those, which have a large number of folding traps and to have a deeper understanding of protein dynamics.

Recent advance in network science opens a lot of possibilities to gain more information from both protein structure networks and conformational/energy networks:

- A systematic comparison and analysis of proper link weights (instead of cut-off distances and un-weighted links) and network building rules (networks of selected key atoms, or of the weighted sum of amino acid side chain atomic coordinates) is a task of the future. Re-analysis of small-worldness in a weighted network may give novel surprises.
- A more refined analysis of the hierarchical and overlapping structure [5,6] of protein structure network modules still holds a lot of surprises in the identification of key protein residues governing enzyme activity, allosteric regulation, function of protein motors, signal transduction and protein stability.
- Modular analysis will also lead to novel centrality-measures going beyond the concept of local centrality (hubs) and global centrality (central residues in the sense of closeness or inverse geodesic length). Centrality indices taking into account weights and all levels of topological structure should be developed and used to identify key protein residues (modular centers, inter-modular bridges and elements of multiple overlapping regions) in a graded manner.
- The introduction of weighted and directed links as well as a systematic hierarchical modular analysis of the conformational/energy networks may solve the long-standing problem of the incomprehensibility of rugged energy landscapes.
- As a later development the introduction of non-paired interactions (hypergraphs like at the early work of Finkelstein and Roytberg [65]) may open a way to analyze even more refined details of protein structure and transitions.
- Finally and most importantly, the analysis of the dynamism and evolution [66] of protein structural networks has not been explored so far. Understanding the dynamics of protein structural networks will help us to understand the complexity of protein dynamics by identifying correlated regions of protein structural networks, which may well correspond to correlated motions of these regions. The introduction of 'protein games' [46] will also help us to understand this complex phenomenon. As an initial finding, cooperative protein regions of protein conformational networks revealed by perturbational analysis gave novel evidence for the central arrangement of active centers [67].

We believe that the literature of protein network studies is right before an expansion. This phenomenon is called as 'tipping point' in networks [68] and shows a sudden increase in the applicability of newly developed concepts. We hope we may have contributed a little to this increase with the current review.

**Acknowledgments**
The authors would like to thank Drs. Judit Fidy, László Nyitray, Péter Tompa, members of the LINK-group (www.weaklinks.sote.hu) and the anonymous referee for helpful comments. Work in the authors' laboratory was supported by research grants from the Hungarian National Science Foundation (OTKA T49213), EU (FP6-506850, FP6-016003) and by the Hungarian National Research Initiative (NKFP-1A/056/2004 and KKK-0015/3.0).

**References**
[1] Barabasi, A.L. and Albert, R. (1999) Emergence of scaling in random networks. Science 286, 509–512.
[2] Barabasi, A.L. and Oltvai, Z.N. (2004) Network biology: understanding the cell's functional organization. Nat. Rev. Genet. 5, 101–113.
[3] Boccaletti, S., Latora, V., Moreno, Y., Chavez, M. and Hwang, D.-U. (2006) Complex networks: structure and dynamics. Physics Rep. 424, 175–308.
[4] Csermely, P. (2006) Weak links: a universal key for network diversity and stability, Springer Verlag, Heidelberg.
[5] Palla, G., Derenyi, I., Farkas, T. and Vicsek, T. (2005) Uncovering the overlapping community structure of complex networks in nature and society. Nature 435, 814–818.
[6] Ravasz, R., Somera, A.L., Mongru, D.A., Oltvai, Z.N. and Barabasi, A.L. (2002) Hierarchical organization of modularity in metabolic networks. Science 297, 1551–1555.
[7] Watts, D.J. and Strogatz, S.H. (1998) Collective dynamics of 'small-world' networks. Nature 393, 440–442.
[8] Arita, M. (2004) The metabolic world of *Escherichia coli* is not small. Proc. Natl. Acad. Sci. USA 101, 1543–1547.
[9] Ma, H.W. and Zeng, A.P. (2003) Reconstruction of metabolic networks from genome data and analysis of their global structure for various organisms. Bioinformatics 19, 220–277.

[10] Stumpf, M.P.H., Wiuf, C. and May, R.M. (2005) Subnets of scale-free networks are not scale-free: sampling properties of networks. Proc. Natl. Acad. Sci. USA 102, 4221–4224.
[11] Tanaka, R., Yi, T. M. and Doyle, J. (2005) Some protein interaction data do not exhibit power law statistics; FEBS Lett. 579, 5140–5144.
[12] Artymiuk, P.J., Rice, D.W., Mitchell, E.M. and Willett, P. (1990) Structural resemblance between the families of bacterial signal-transduction proteins and of G proteins revealed by graph theoretical techniques. Protein Eng. 4, 39–43.
[13] Mitchell, E.M., Artymiuk, P.J., Rice, D.W. and Willett, P. (1990) Use of techniques derived from graph theory to compare secondary structure motifs in proteins. J. Mol. Biol. 212, 151–166.
[14] Kannan, N. and Vishveshwara, S. (1999) Identification of side-chain clusters in protein structures by a graph spectral method. J. Mol. Biol. 292, 441–464.
[15] Aftabuddin, M. and Kundu, S. (2006) Weighted an unweighted network of amino acids in a protein. Physica A 396, 895–904.
[16] Vendruscolo, M., Dokholyan, N.V., Paci, E. and Karplus, M. (2002) Small-world view of the amino acids that play a key role in protein folding. Phys. Rev. E 65, 061910.
[17] Atilgan, A.R., Akan, P. and Baysal, C. (2004) Small-world communication of residues and significance for protein dynamics. Biophys. J. 86, 85–94.
[18] Bagler, G. and Sinha, S. (2005) Network properties of protein structures. Physica A 346, 27–33.
[19] Alves, N.A. and Martinez, A.S. (2007) Inferring topological features of proteins from amino acid residue networks. Physica A 375, 336–344.
[20] Brinda, K.V. and Vishveshwara, S. (2005) A network representation of protein structures: implications for protein stability. Biophys. J. 89, 4159–4170.
[21] Dokholyan, N.V., Li, L., Ding, F. and Shakhnovich, E.I. (2002) Topological determinants of protein folding. Proc. Natl. Acad. Sci. USA 99, 8637–8641.
[22] Del Sol, A., Fujihashi, H. and Nussinov, R. (2006) Residues crucial for maintaining short paths in network communication mediate signalling in proteins. Molec. Systems. Biol. 2006.0019
[23] Amitai, G., Shemesh, A., Sitbon, E., Shklar, M., Netanely, D., Venger, I. and Pietrokovski, S. (2004) Network analysis of protein structures identifies functional residues. J. Mol. Biol. 344, 1135–1146.
[24] Bikadi, Z., Demko, L. and Hazai, E. (2007) Functional and structural characterization of a protein based on analysis of its hydrogen bonding network by hydrogen bonding plot. Arch. Biochem. Biophys. in press
[25] Aftabuddin, M. and Kundu, S. (2007) Hydrophobic, hydrophilic and charged amino acids' networks within protein. Biophys. J. in press
[26] Greene, L.H. and Higman, V.A. (2003) Uncovering network systems within protein structures. J. Mol. Biol. 334, 781–791.
[27] Scala, A., Nunes Amaral, L.A. and Barthelemy, M. (2001) Small-world networks and the conformation space of a short lattice polymer chain. Europhys. Lett. 55, 594–600.
[28] Wangikar, P.P., Tendulkar, A.V., Ramya, S., Mali, D.N. and Sarawagi, S. (2003) Functional sites in protein families uncovered via an objective and automated graph theoretic approach. J. Mol. Biol. 326, 955–978.
[29] Balog, E., Laberge. M. and Fidy, J. (2007) The influence of interdomain interactions on the intradomain motions in yeast phosphoglycerate kinase: a molecular dynamics study. Biophys. J. 92, 1709–1716.
[30] Hayward, S. and Berendsen, H.J.C. (1998) Systematic analysis of domain motions in proteins from conformational change: new results on citrate synthase and T4 lysozyme. Proteins 30, 144–154.
[31] Dosztanyi, Z., Magyar, C., Tusnady, G.E. and Simon, I. (2003) SCide: identification of stabilization centers in proteins. Bioinformatics 19, 899–900.
[32] Koonin, E.V., Wolf, Y.I. and Karev, G.P. (2002) The structure of the protein universe and genome evolution. Nature 420, 218–223.
[33] Sickmeier, M., Hamilton, J.A., LeGall, T., Vacic, V., Cortese, M.S., Tantos, A., Szabo, B., Tompa, P., Chen, J., Uversky, V.N., Obradovic, Z. and Dunker, A.K. (2007) DisProt: the Database of Disordered Proteins. Nucleic Acids Res. 35, D786–D793.
[34] Vanderkooi, J.M., Kaposi, A. and Fidy, J. (1993) Protein conformation monitored by energy-selective optical spectroscopy. Trends Biochem. Sci. 18, 71–76.
[35] Ansari, A., Berendzen, J., Bowne, S.F., Frauenfelder, H., Iben, I.E.T., Sauke, T.B., Shyamsunder, E. and Young, R.D. (1985) Protein states and proteinquakes. Proc. Natl. Acad. Sci. USA 82, 5000–5004.
[36] Dewey, T.G. and Bann, J.G. (1992) Protein dynamics and 1/f noise. Biophys. J. 63, 594–598.
[37] Flomenbom, O., Velonia, K., Loos, D., Masuo, S., Cotlet, M., Engelborghs, Y., Hofkens, J., Rowan, A.E., Nolte, R.J.M., van der Auweraer, M. and de Schryver, F.C. (2005) Stretched exponential decay and correlations in the catalytic activity of fluctuating single lipase molecules. Proc. Natl. Acad. Sci. USA 102, 2368–2372.
[38] Metzler, R., Klafter, J., Jortner, J. and Volk, M. (1998) Multiple time scales for dispersive kinetics in early events of peptide folding. Chem. Phys. Lett. 293, 477–484.

[39] Okazaki, K., Koga, N., Takada, S., Onuchic, J.N. and Wolynes, P.G. (2006) Multiple-basin energy landscapes for large-amplitude conformational motions of proteins: Structure-based molecular dynamics simulations. Proc. Natl. Acad. Sci. USA 103, 11844–11849.
[40] Agarwal, P.K., Billeter, S.R., Rajagopalan, P.T.R., Benkovic, S.J. and Hammes-Schiffer, S. (2002) Network of coupled promoting motions in enzyme catalysis. Proc. Natl. Acad. Sci. USA 99, 2494–2499.
[41] Zheng, W., Brooks, B.R. and Thirumalai, D. (2006) Low-frequency normal modes that describe allosteric transitions in biological nanomachines are robust to sequence variations. Proc. Natl. Acad. Sci. USA 103, 7664–7669.
[42] Esienmesser, E.Z., Millet, O., Labeikovsky, W., Korzhnev, D.M., Wolf-Watz, M., Bosco, D.A., Skalicky, J.J., Kay, L.E. and Kern, D. (2005) Intrinsic dynamics of an enzyme underlies catalysis. Nature 438, 117–121.
[43] Jacobs, D.J., Rader, A.J., Kuhn, L.A. and Thorpe, M.F. (2001) Protein flexibility predictions using graph theory. Proteins 44, 150–165.
[44] Barron, L.D., Hecht, L. and Wilson, G. (1997) The lubricant of life: a proposal that solvent water promotes extremely fast conformational fluctuations in mobile heteropolypeptide structure. Biochemistry 36, 13143–13147.
[45] Csermely, P. (2001) Water and cellular folding processes. Cell. Mol. Biol. 47, 791–800.
[46] Kovacs, I.A., Szalay, M.S. and Csermely, P. (2005) Water and molecular chaperones act as weak links of protein folding networks: energy landscape and punctuated equilibrium changes point towards a game theory of proteins. FEBS Lett. 579, 2254–2260.
[47] Levy, Y. and Onuchic, J.N. (2006) Water mediation in protein folding and molecular recognition. Annu. Rev. Biophys. Biomol. Struct. 35, 389–415.
[48] Papoian, G.A., Ulander, J., Eastwood, M.P., Luthey-Schulten, Z. and Wolynes, P.G. (2004) Water in protein structure prediction. Proc. Natl. Acad. Sci. USA 101, 3352–3357.
[49] Amisha-Kamal, J.K., Zhao, L. and Zewail, A.H. (2004) Ultrafast hydration dynamics in protein unfolding: Human serum albumin. Proc. Natl. Acad. Sci. USA 101, 13411–13416.
[50] Oleinikova, A., Brovchenko, I., Smolin, N., Krukau, A., Geiger, A. and Winter, R. (2005) The percolation transition of hydration water: from planar hydrophilic surfaces to proteins. Phys. Rev. Lett. 95, 247802.
[51] Kurkal, V., Daniel, R.M., Finney, J.L., Tehei, M., Dunn, R.V. and Smith, J.C. (2005) Enzyme activity and flexibility at very low hydration. Biophys. J. 89, 1282–1287.
[52] Fenimore, P.W., Frauenfelder, H., McMahon, B.H. and Parak, F.G. (2002) Slaving: solvent fluctuations dominate protein dynamics and functions. Proc. Natl. Acad. Sci. USA 99, 16047–16051.
[53] Klibanov, A.M. (1995) What is remembered and why? Nature 374, 596.
[54] Csermely, P. (2004) Strong links are important, but weak links stabilize them. Trends Biochem. Sci. 29, 331–334.
[55] Doye, J.P.K. (2002) The network topology of a potential energy landscape: A static scale-free network. Phys. Rev. Lett. 88, 238701.
[56] Doye, J.P.K. and Massen, C.P. (2005) Characterizing the network topology of the energy landscapes of atomic clusters. J. Chem. Phys. 122, 084105.
[57] Gfeller, D., De Los Rios, P., Caflisch, A. and Rao, F. (2007) Complex network analysis of free-energy landscapes. Proc. Natl. Acad. Sci. USA 104, 1817–1822.
[58] Doye, J.P.K. and Massen, C.P. (2006) Energy landscapes, scale-free networks and Apollonian packings. In: Complexity, Metastability and Nonextensivity. 31st Workshop of the International School of Solid State Physics (Eds. C. Beck, G. Benedek, A. Rapisarda and C. Tsallis) World Scientific, Singapore.
[59] Rylance, G.J., Johnston, R.L., Matsunaga, Y., Li, C.B., Baba, A. and Komatsuzaki, T. (2006) Topographical complexity of multidimensional energy landscapes. Proc. Natl. Acad. Sci. USA 103, 18551–18555.
[60] Herenyi, L., Szigeti, K., Fidy, J., Temesvari, T., Schlichter, J. and Friedrich, J. (2004) Aging dynamics in globular proteins: summary and analysis of experimental results and simulation by a modified trap model. Eur. Biophys. J. 33, 68–75.
[61] Osvath, S., Herenyi, L., Zavodszky, P., Fidy, J. and Kohler, G. (2006) Hierarchic finite level energy landscape model: to describe the refolding kinetics of phosphoglycerate kinase. J. Biol. Chem. 281, 24375–24380.
[62] Yang, H., Luo, G., Karnchanaphanurach, P., Louie, T.-M., Rech, I., Cova, S., Xun, L. and Xie, X.S. (2003) Protein conformational dynamics probed by single-molecule electron transfer. Science 302, 262–266.
[63] Rao, F. and Caflisch, A. (2004) The protein folding network. J. Mol. Biol. 342, 299–306.
[64] Samudrala. R. and Moult, J. (1998) A graph-theoretic algorithm for comparative modeling of protein structure. J. Mol. Biol. 279, 287–302.
[65] Finkelstein, A. V. and Roytberg, M. A. (1993) Computation of biopolymers: a general approach to different problems. Biosystems 30, 1–19.
[66] Leskovec, J., Kleinberg, J. and Faloutsos, C. (2006) Laws of graph evolution: densification and shrinking diameters. ACM Transact. Knowledge Discov. 1, 1–40.
[67] Liu, T., Whitten, S.T. and Hilser, V.J. (2007) Functional residues serve a dominant role in mediating the cooperativity of the protein ensemble. Proc. Natl. Acad. Sci. USA 104, 4347–4352.
[68] Gladwell, M. (2000) The tipping point: How little things can make a big difference. Little Brown, Boston MA USA.